\documentstyle[preprint,floats,aps]{revtex}

\newcommand\lapprox{\mathrel{\mathop
  {\hbox{\lower0.5ex\hbox{$\sim$}\kern-0.8em\lower-0.7ex\hbox{$<$}}}}}
\newcommand\gapprox{\mathrel{\mathop
  {\hbox{\lower0.5ex\hbox{$\sim$}\kern-0.8em\lower-0.7ex\hbox{$>$}}}}}

\begin{document}
\preprint{MPI-Ph/93-81}
\title{Experimental and Astrophysical Status
       of Neutrino Masses\footnote{Lecture given at the Les Houches
       School ``Cosmology and Large Scale Structure'', August 1993.
       To appear in the proceedings.}}

\author{Georg Raffelt\\
Max-Planck-Institut f\"ur Physik \\
F\"ohringer Ring 6, 80505 M\"unchen, Germany}

\maketitle

\begin{abstract}
\noindent The experimental and astrophysical status of neutrino masses
is reviewed with an emphasis on the cosmologically interesting regime.
\end{abstract}


\section{Introduction}
It is widely believed that a large fraction of the cosmic dark matter consists
of some sort of weakly interacting particles. Among the known forms of matter
neutrinos are the only possible candidates, although a variety of hypothetical
particles have been considered which may well play this role.  Neutrinos would
fall into the category of hot dark matter which is disfavored by our current
understanding of cosmic structure formation. However, recent discussions of
mixed hot and cold dark matter scenarios have revived an interest in neutrinos
as a substantial matter component of the universe, and so it remains of
interest to cosmologists what is empirically known about the masses of these
elusive particles.

As discussed in any textbook on cosmology, the neutrino contribution to the
cosmic mass density is $\Omega_\nu h^2=\sum m_{\nu}/93\,{\rm eV}$.  From the
CERN measurements of the $Z^\circ$ decay width we know that there is no fourth
family with a mass below $m_Z/2\approx 46\,{\rm GeV}$ \cite{ParticleData} so
that one may safely limit the discussion to the three known flavors $\nu_e$,
$\nu_\mu$, and $\nu_\tau$. Current values for the cosmological parameters seem
to indicate $\Omega h^2\lapprox 0.4$ so that $m_\nu\lapprox 40\,{\rm eV}$ for
all flavors.  This bound could only be evaded if neutrinos decayed much faster
than is allowed within the well-established standard model of particle
interactions.  Therefore, it is crudely the decade $3-30\,\rm eV$ of neutrino
masses which is of cosmological interest.


\section{Direct Bounds}
\subsection{Tritium Beta Decay}
Neutrinos were first ``discovered'' in weak nuclear decays of the form
$(A,Z)\to\break(A,Z+1) e^- \overline\nu_e$ where the continuous energy
spectrum of
the electrons reveals the emission of another particle that carries away the
remainder of the available energy~\cite{Winter}. The minimum amount of energy
taken by the neutrino is the equivalent of its mass so that the upper endpoint
of the electron spectrum is a sensitive measure for $m_{\nu_e}$.  Actually,
the most sensitive probe is the {\it shape\/} of the electron spectrum just
below its endpoint, not the value of the endpoint itself. The best constraints
are based on the tritium decay
${}^3{\rm H}\to{}^3{\rm He}^+\,e^-\overline\nu_e$ with a maximum amount of
kinetic energy for the electron of $Q=18.6\,{\rm keV}$. This unusually small
$Q$-value ensures that a relatively large fraction of the electron counts
appear near the endpoint~\cite{Winter}.

\begin{table}
\caption{Summary of tritium $\beta$ decay experiments}
\begin{tabular}
{llccc}
Experiment&Year&\vrule depth 3ex height 4ex width 0pt%
$\displaystyle{{m_{\nu_e}^2\pm\sigma_{\rm stat}
\pm\sigma_{\rm syst}}\atop{[{\rm eV}^2]}}$%
&\vrule depth 3ex height 4ex width 0pt$\displaystyle{{\rm 95\%~CL~Upper}\atop
{\rm Limit~}m_{\nu_e}[{\rm eV]}}$&Ref.\\
\hline
Los Alamos&1991&$-147\pm68\pm41$&9.3&\cite{Robertson}\\
Tokyo     &1991&$-65\pm85\pm65$&13&\cite{Kawakami}\\
Z\"urich  &1992&$-24\pm48\pm61$&11&\cite{Holzschuh}\\
Livermore &1992&$-72\pm41\pm30$&8&\cite{Stoeffl}\\
Mainz     &1992&$-39\pm34\pm15$&7.2&\cite{Weinheimer}\\
\end{tabular}
\end{table}

In Table~I we quote the results from several recent experiments which
had been
motivated by the Moscow (1987) claim of $17\,{\rm eV}<m_{\nu_e}<40\,{\rm eV}$
\cite{Boris}.  This range is clearly incompatible with the more recent data
which, however, find {\it negative\/} masses squared with a world average of
$m^2_{\nu_e}=-(59\pm26)\,{\rm eV}^2$ (Wilkerson in~\cite{Neutrino92}). This
result means that the endpoint spectra tend to be slightly deformed in the
opposite direction from what a neutrino mass would do. While this effect is
perhaps due to a statistical fluctuation one may be worried that there are
still significant systematic uncertainties. Nominally, the combined results of
Table~I imply a 95\%~CL upper limit of $5\,\rm eV$.

\subsection{Mu and Tau Neutrinos}
To constrain $m_{\nu_\mu}$ one may measure the muon momentum from the decay of
stopped pions, $\pi^+\to\mu^+\nu_\mu$, so that
$m_{\nu_\mu}^2=m_{\pi^+}^2+m_\mu^2-2m_{\pi^+}(m_\mu^2+p_\mu^2)^{1/2}$.
However, a recent measurement~\cite{Daum} implies a negative mass squared of
$m_{\nu_\mu}^2=-(0.154\pm0.045)\,{\rm MeV}^2$, probably due to large
systematic uncertainties in the determination of $m_{\pi^+}$. Hence, the
usually quoted bound of $m_{\nu_\mu}<0.27\,{\rm MeV}$ does not really apply.
An older experiment studied the in-flight decay of pions with a result
$m_{\nu_\mu}^2=-(0.14\pm0.20)\,{\rm MeV}^2$, largely independent of the pion
mass~\cite{Anderhub}. This implies a 90\%~CL upper limit of
$m_{\nu_\mu}<0.50\,{\rm MeV}$.

For $\nu_\tau$ the best bounds also come from limits on missing energy in
certain reactions, the only form in which $\nu_\tau$ has ever been
``observed''. The ARGUS collaboration studied the decay
$\tau^-\to3\pi^-2\pi^+\nu_\tau$ with a total of 20 events with good energy
determinations for all five pions, leading to $m_{\nu_\tau}<31\,{\rm MeV}$ at
95\%~CL~\cite{ARGUS}.  A similar experiment by the CLEO collaboration based on
a much larger data sample recently gave $m_{\nu_\tau}<32.6\,{\rm MeV}$ at
95\%~CL~\cite{CLEO}.

Evidently it will not be possible to improve these methods to become sensitive
to $\nu_\mu$ or $\nu_\tau$ masses in the cosmologically interesting range.
The higher families of charged leptons and quarks are each much more massive
than the previous family so that one would also expect that
$m_{\nu_e}\ll m_{\nu_\mu}\ll m_{\nu_\tau}$, even though it is possible that
neutrino masses are arranged in a different order. Any hope of finding
cosmological neutrino masses by direct kinematic methods is extremely remote.

\subsection{Supernova Neutrinos}
Other direct methods which assume nothing about neutrinos except their mass
involve astrophysics, most prominently among them the cosmological constraint.
Furthermore, the observed neutrino signal from SN 1987A allowed one to limit
$m_{\nu_e}$ from the absence of an anomalous signal dispersion that would
occur because the velocity of massive neutrinos depends on their energy as
$v/c=(1-m_\nu^2/E_\nu^2)^{1/2}$. A detailed analysis~\cite{Loredo} yields
$m_{\nu_e}\lapprox23\,{\rm eV}$, less restrictive than the tritium bounds.

However, the situation would change if we were to observe a galactic SN with,
for example, the Superkamiokande detector which is currently under
construction. In such a water \v Cerenkov detector neutrinos are measured
through $\overline\nu_ep\to n e^+$ (positron approximately isotropic) and by
$\nu e^-\to e^-\nu$ (electron forward peaked).  The latter reaction, being of
the neutral-current type, is sensitive to neutrinos and antineutrinos of all
flavors and so, it can serve to monitor the $\nu_\tau$ flux while the former
reaction, which yields a much stronger signal because the cross section is
much larger, allows to monitor the thermal evolution of the SN core. The low
laboratory constraints on $m_{\nu_e}$ reveal that we would be able to follow
directly the time structure of neutrino emission without excessive $\nu_e$
dispersion effects. A dispersion of the $\nu_\tau$ signal due to a mass term
could then possibly be identified down to $m_{\nu_\tau}\gapprox25\,{\rm
eV}$~\cite{Krauss}.

\newpage
\section{Indirect Searches}
\subsection{Oscillation Experiments}
Apart from waiting for a galactic SN, the only realistic chance to detect
small neutrino masses is to use indirect methods, notably those based on
neutrino oscillations. In analogy to the quarks one expects that the weak
interaction eigenstates $\nu_\ell$, $\ell=e,\mu,\tau$, are linear combinations
of the mass eigenstates $\nu_i$, $i=1,2,3$, by virtue of a unitary matrix,
$\nu_\ell=\sum_{j=1}^3 U_{\ell j}\nu_j$. If a neutrino is produced by a
charged-current reaction it will be in a definite flavor eigenstate and thus a
mixture of mass eigenstates.  For a given energy $E_\nu$ the momentum of one
mass component $j$ is $p_j=(E_\nu^2-m_j^2)^{1/2}\approx E_\nu-m_j^2/2E_\nu$.
Because the phase of a given component as a function of distance $r$ evolves
as $e^{i p_j r}$, the components develop relative phase differences and so, we
find a flavor composition different from the original state as a function of
distance from the source. The phase difference is proportional to the
differences of $m_j^2$ so that oscillation experiments are sensitive to the
quantity $\Delta m^2$ of the mixing flavors. Moreover, they are sensitive to
the parameters $U_{\ell j}$ which are usually parametrized in terms of mixing
angles.\footnote{The experiments are usually analyzed in terms of two-flavor
mixing whence the mixing matrix may be represented as
$U=\pmatrix{\cos\theta&\sin\theta\cr-\sin\theta&\cos\theta\cr}$.}

No experimental evidence for oscillations has been found, although a host of
experiments at reactors and accelerators has excluded various regions in the
$\Delta m^2$-$\sin^22\theta$-plane for various combinations of
flavors~\cite{ParticleData,Winter,Neutrino90}. Cosmologists are interested in
the range for $\Delta m^2$ of about $10-10^3\,\rm eV^2$ if we ignore the
possibility of almost degenerate mass eigenstates. For
$\Delta m^2\gapprox 10\,\rm eV^2$ one finds $\sin^22\theta\lapprox0.1$ for
$\nu_e\leftrightarrow\nu_\tau$ and $\sin^22\theta\lapprox0.004$ for
$\nu_\mu\leftrightarrow\nu_e$ or $\nu_\mu\leftrightarrow\nu_\tau$, almost
independently of $\Delta m^2$.

It is very encouraging that at CERN two experiments (CHARM and NOMAD) are
currently in preparation whose goal is to improve the sensitivity to
$\nu_\mu\to\nu_\tau$ oscillations (DiLella in \cite{Neutrino92}).
For $\Delta m^2\gapprox 10\,\rm eV^2$ oscillations should be detectable down
to $\sin^22\theta\gapprox 1{\times}10^{-4}$. Both experiments are expected to
start taking data in the spring of 1994.  Therefore, it is quite conceivable
that cosmologically interesting neutrino masses will be detected at CERN
within the next few years.

\subsection{Solar Neutrinos}
The first positive indication for neutrino oscillations is provided by the
solar neutrino measurements which show a deficit in the high-energy part of
the spectrum relative to the calculated flux based on detailed solar models.
The data can be consistently interpreted by two-flavor oscillations with two
solutions:
(a)~$\Delta m^2=(3-12)\,\rm meV^2$ and $\sin^22\theta=(0.4-1.5)\times10^{-2}$;
(b)~$\Delta m^2=(3-40)\,\rm meV^2$ and $\sin^22\theta=0.5-0.9$ (e.g.~Bludman
et al.\ in~\cite{Neutrino92}; for a detailed discussion of solar neutrinos see
M.~Spiro in these proceedings).

If future experimental and theoretical works confirm this tentative
interpretation, and if in the Sun the $\nu_e$ oscillate into $\nu_\mu$, we are
left with the possibility of a cosmologically interesting $m_{\nu_\tau}$. With
non-degenerate masses ($m_{\nu_e}\ll m_{\nu_\mu}$) the Sun would indicate
$m_{\nu_\mu}=2-6\,\rm meV$ so that $m_{\nu_\tau}$ in the cosmological range
$3-30\,\rm eV$ requires
$10^{-4}\lapprox m_{\nu_\mu}/m_{\nu_\tau}\lapprox10^{-3}$. This mass ratio may
be reasonable in view of so-called see-saw models for neutrino masses which
predict $m_{\nu_e}:m_{\nu_\mu}:m_{\nu_\tau}=m_e^2:m_\mu^2:m_\tau^2$ or
$m_u^2:m_c^2:m_t^2$ (up, charm, and top quark masses)~\cite{Winter}. Note that
$m_\mu^2/m_\tau^2=3.5{\times}10^{-3}$ and
$m_c^2/m_t^2\approx (0.3-3){\times}10^{-4}$,
although the top quark has not yet been directly observed.

\subsection{Atmospheric Neutrinos}
A much more controversial indication for neutrino oscillations comes from the
observation of neutrinos produced by cosmic rays. High-energy protons produce
showers in the atmosphere which contain large numbers of pions. Charged pions
dominantly decay via $\pi\to\mu\nu_\mu$ and subsequently
$\mu\to e\nu_e\nu_\mu$ where all symbols stand for particles as well as
antiparticles. Therefore, one expects an atmospheric (anti) neutrino flux with
the rather precise and model-independent flavor composition
$\nu_e:\nu_\mu=1:2$.  However, in the low-energy part of the spectrum the
underground water \v Cerenkov detectors IMB and Kamiokande measured a deficit
of $\nu_\mu$ by almost a factor of two, i.e., they found
$\nu_e:\nu_\mu\approx1:1$. Other detectors based on iron calorimeters
(Fr\'ejus, NUSEX) did not find this discrepancy, and IMB and Kamiokande do not
see a discrepancy in the high-energy flux of upward going muons (for a review
see Totsuka in~\cite{Neutrino92} and for a more recent result~\cite{Becker}).

Still, there may remain a region of $(\Delta m^2,\sin^22\theta)$ around
$(10^{-2}\,{\rm eV}^2,0.5)$ which allows for a consistent interpretation in
terms of neutrino oscillations. Notably, an interpretation in terms of
$\nu_\mu$-$\nu_\tau$ oscillations is not in conflict with the solar neutrino
result.  However, it would not allow for a cosmologically significant
$m_{\nu_\tau}$ unless one appeals to almost degenerate neutrino masses.

The identification procedure for $\mu$'s and $e$'s depends on subtle
differences in their pattern of \v Cerenkov light. Currently an effort is
under way to build a prototype detector at KEK in order to calibrate that
procedure by means of $e$ and $\mu$ beams (Totsuka in~\cite{Neutrino92}).
Therefore, we may hope that the atmospheric neutrino problem will be cleared
up within the next few years.

\subsection{R-Process in Supernovae}
It is widely believed that the rapid neutron capture process ($r$-process) for
the formation of heavy nuclei occurs in the neutron rich environment of a SN.
The $\nu_{\mu}$ or $\nu_\tau$ neutrinos emitted from a SN core are more
energetic than the $\nu_e$'s so that a possible swapping of these flavors by
oscillations would result in more energetic $\nu_e$'s and thus in enhanced
$\nu_e+n\to p+e^-$ conversions, depleting the material of neutrons and thus
suppressing the heavy element synthesis. Because of the high density of the
material just outside the neutrino sphere, resonant neutrino oscillations can
occur for $m_\nu$'s in the cosmologically interesting regime. Recent
investigations~\cite{Qian} showed that the $r$-process would be suppressed for
$\Delta m^2$ corresponding to $m_{\nu_\mu}$ or $m_{\nu_\tau}$ in the range
$1-100\,\rm eV$ and $\sin^22\theta>10^{-5}$. Therefore, either the neutrino
parameters do not lie in this range, or SN are not the site of the
$r$-process, with no known alternative.

\subsection{Neutrinoless Double Beta Decay}
There are several isotopes which can decay only by the simultaneous conversion
of two neutrons, $(A,Z)\to(A,Z+2)\,2e^-\,2\overline\nu_e$. Recently it has
become possible to observe the electron spectra from such second-order weak
decays directly~\cite{Neutrino92}. For example, the decay
${}^{76}{\rm Ge}\to{}^{76}{\rm Se}\,2e^-\,2\overline\nu_e$ is found to have a
half life of $(1.43\pm0.04_{\rm stat}\pm0.13_{\rm syst})\times10^{21}\,\rm yr$
\cite{Beck}. Of much greater interest, however, is the possibility of a
neutrinoless decay mode $(A,Z)\to(A,Z+2)\,2e^-$ which would violate lepton
number by two units. In a measurement of the combined energy spectrum of both
electrons the $0\nu$ mode would show up as a peak at the endpoint.

Lepton number would be violated if neutrinos were their own antiparticles,
so-called Majorana neutrinos as opposed to Dirac ones with four distinct
states like electrons ($e^\pm$, spin up and down each).  Because weak
interactions violate parity maximally, the two right-handed Dirac neutrino
states would be inert, and there is no practical distinction between Dirac and
Majorana neutrinos unless they have a mass.  Loosely speaking, the $0\nu$
decay mode is then possible because one of the emitted neutrinos is
re-absorbed as an anti-neutrino. The amplitude for this process is
proportional to $m_{\nu_e,{\rm Majorana}}$ and so, the rate for the $0\nu$
decay mode is proportional to $m_{\nu_e,{\rm Majorana}}^2$.  Current upper
bounds are around $1\,\rm eV$, the exact value depends on nuclear matrix
elements which are not precisely known.

With neutrino mixing the other flavors also contribute so that the bound is
really on the quantity
$\langle m_\nu\rangle\equiv\sum_j \lambda_j\vert U_{ej}\vert^2 m_j$ where
$\lambda_j$ is a CP phase equal to $\pm1$, and the sum is to be extended over
all two-component Majorana neutrinos.\footnote{In this language a
four-component Dirac neutrino consists of two degenerate two-component
Majorana ones with $\lambda=+1$ and $-1$ so that their contributions cancel
exactly, reproducing the absence of lepton number violation for Dirac
neutrinos.} If we take $m_{\nu_\tau}=30\,\rm eV$, for example, and 0.16 for
$\vert U_{e,3}\vert$ we may have a contribution as large as $0.8\,\rm eV$ from
the $\nu_\tau$.

Interestingly, there have been recent reports of very preliminary indications
of a small number of excess counts at the endpoints of the $2\beta$ spectra in
the Moscow-Heidelberg ${}^{76}{\rm Ge}$ \cite{Piepke} and the
Milano ${}^{130}{\rm Te}$ experiments \cite{Garcia}. The current low
statistical significance of perhaps $1.5-2\,\sigma$ requires much more data
before any conclusions regarding $\langle m_\nu\rangle$ can be drawn.


\section{Summary}
Current direct bounds on neutrino masses are essentially insignificant with
regard to detecting or excluding cosmologically relevant neutrino masses.
However, several current efforts employing indirect methods based on neutrino
mixing may yet turn up definitive evidence for non-vanishing $m_\nu$'s. The
frontrunners are solar neutrinos with several new experiments coming on line,
and running ones being calibrated, within the next few years, and the CERN
$\nu_\mu$-$\nu_\tau$ oscillation experiments with results expected in two to
three years. As a dark horse it is still possible that the atmospheric
neutrino problem can be attributed to neutrino oscillations, but a
calibration of the \v Cerenkov detectors must be awaited.  It will be
exciting to see if more data confirm the recent endpoint excess counts in
$2\beta$ decay experiments that would signify the existence of Majorana
neutrino masses. In summary, the 1990's promise a rich harvest of new and
perhaps definitive results with regard to cosmological neutrino masses.


\end{document}